\documentstyle[12pt,amsfonts]{article}

\def\half{\textstyle{\frac{1}{2}}}

\def\H{{\cal H}}
\def\D{{\cal D}}

\def\l{\lambda}
\def\E{{\rm I}\hskip-.2em{\rm E}}
\def\ra{\rightarrow}
\def\tint{{\textstyle\int}}
\def\hg{{\hat g}}
\def\hp{{\hat\pi}}

\def\s{\hskip.08em}
\def\d{\partial}
\def\o{\overline}
\def\a{\alpha}
\def\b{\begin{eqnarray*}}  
\def\e{\end{eqnarray*}}    
\def\bn{\begin{eqnarray}}  
\def\en{\end{eqnarray}}   
\def\<{\langle}
\def\>{\rangle}

\def\no{\nonumber}
\let\mathbb\cal

\def\{{\lbrace}
\def\}{\rbrace}
\begin{document}
\title{Overview of Affine Quantum Gravity}
\author{John R.~Klauder\\
Department of Physics and Department of Mathematics\\
University of Florida\\
Gainesville, FL 32611}
\date{}    
\maketitle
\begin{abstract}
The basic principles of Affine Quantum Gravity are presented in a pedagogical style with a limited number of equations.
\end{abstract}
\section{Introduction}

No one should doubt the fact that quantum gravity is a difficult problem to
deal with. String theory and loop quantum gravity are the standard views
on the subject, and they both comprise major efforts. However, in recent years  
a more modest program called {\it Affine
Quantum Gravity} has begun. The goal of this paper is to summarize motivations and present some of the
developments in that program. In the present section, we offer a broad overview in words.
In the next section, we add a few equations to these words to illustrate what
is done in practice. Our overall story is based principally on the following
references: \cite{kl1,kl2,kl3,kl4}. 

The approach adopted with affine quantum gravity is designed to stay as
close to the classical theory of gravity as possible. This approach enables
one to clearly see how the correspondence limit may arise and how the given
quantum theory can pass back to the known classical theory as $\hbar\ra0$,
or as suitable quasi-classical quantum states are considered. In this approach we have chosen fields to promote to basic kinematical operators that
have a direct appearance in the classical theory. As usual in quantization, other
operators are constructed from the algebra of the kinematical operators. In particular, 
the other operators of initial interest are generators of constraints, which for gravity
are the three diffeomorphism constraints and one Hamiltonian (or temporal)
constraint, for a $3+1$ split of spacetime. Classically these constraint fields
have a set of mutual Poisson brackets that form an open first-class algebra. 
That is, the brackets have the appearance of a Lie algebra save for the fact that
the structure constants are replaced in part by structure functions. Upon
quantization, the gravitational constraint fields become locally self-adjoint field operators,
and the corresponding set of commutator brackets does {\it not} describe a set of
first-class constraints, but rather a set of partially second-class constraints due to the
nature of the structure functions involved;
cf. Eq. (\ref{secclass}) below.\footnote{This situation
is in marked contrast to the case of Yang Mills fields for which the Poisson 
brackets form a closed set of first-class constraints, namely having the appearance of a Lie algebra with the structure constants being strictly numerical. On quantization, the 
constraints of Yang Mills theories lead to self-adjoint operator fields that have a set of commutator brackets which forms a quantum set of first-class constraint operators. Many of the 
significant distinctions between Yang Mills theories and gravity were already apparent to Utiyama
is his pioneering work of 1956 \cite{uti}.} Opinions differ as to what to do about the
partial second-class nature of the quantum constraints, and most workers choose to
modify the initial theory in such a way that the quantum constraints actually become
first class. This choice reflects the fact that most techniques to deal with quantum
constraints are designed to work for first-class systems and involve significant modification
to be applied to second-class systems. A relatively new procedure to deal with the 
quantization of systems with constraints is known as the {\it projection operator method}
and one of its strengths is the fact that it deals with first- and second-class quantum constraints by the same procedure and on an equal footing \cite{KORIG, UNIV, SCHAL}. We choose to accept
that gravity is qualitatively different from most other gauge theories, and that there
is physics contained in the fact that the constraints become partially second-class on
quantization. Thus, the projection operator method appears to be ideally suited to be applied to the case of quantum gravity. 

We note that operators are fine in principle but
they are sometimes rather difficult to deal with in practice. That is why so many workers
in various fields of physics, but particularly those who deal with quantum field theory,
have a fondness for functional integral formulations of the quantization process. Thus
it is of some interest that the quantization procedures for affine quantum gravity outlined
above all admit useful functional integral realizations. 

With functional integrals in mind, we note that we expect to deal with the nonrenormalizability of traditional quantum gravity on the basis of a hard core picture of its nonlinear interaction. The interpretation of nonrenormalizable interactions in this fashion has been supported by soluble nonrenormalizable models \cite{kbook}, and it offers a natural viewpoint for gravity as well \cite{nongrav}. 

Let us next turn to some basic equations relevant for affine quantum gravity that serve to illuminate the previous remarks.
\section{Basic Relations}

\subsubsection*{Metric positivity}
A fundamental property of affine quantum gravity is the {\it strict positivity of the spacial metric}. For the classical
metric $g_{ab}(x)$, this property means that for any nonvanishing set $\{u^a\}$ of real numbers and any 
nonvanishing, nonnegative test function, $f(x)\ge0$, that
 \bn  \tint f(x)\s u^a g_{ab}(x)\s u^b\s d^3\!x>0\;.  \en
For the quantum operator associated to the spacial metric, a similar criterion is adopted, namely that 
 \bn  \tint f(x)\s u^a\s\hg_{ab}(x)\s u^b\s d^3\!x>0\;,  \en
where $\hg_{ab}(x)$ denotes the $3\times3$ operator metric field.

\subsubsection*{Affine commutation relations}
Canonical commutation relations are not compatible with the requirement of metric
positivity. This fact holds because the canonical momentum acts to translate the spectrum of the metric tensor
and such a translation is incompatible with metric positivity. Thus it is necessary to find a 
suitable but distinctly alternative set of commutation relations. A suitable alternative 
that has the virtue of preserving the spectrum of a positive metric operator is
not hard to find.

The initial step involves replacing the classical ADM canonical momentum $\pi^{ab}(x)$
with the classical mixed-index momentum $\pi^a_b(x)\equiv \pi^{ac}(x)g_{cb}(x)$. We refer
to $\pi^a_b(x)$ as the ``momentric" tensor being a combination of the canonical {\it momen}tum
and the canonical me{\it tric}. (The author has also referred to $\pi^a_b(x)$ as the ``scale" field, but that name was always regarded as merely a place holder for something better.)
Besides the metric being promoted to an operator, $\hg_{ab}(x)$, we also promote the 
classical momentric tensor to an operator field, $\hp^a_b(x)$; this pair of operators forms the basic kinematical affine operator fields, and all operators of interest are given as functions of this fundamental pair. The basic kinematical operators are chosen so that they satisfy the
following set of {\it affine commutation relations} (in units where $\hbar=1$, which are used throughout unless stated otherwise):
\bn
  &&\hskip.2cm[\hp^a_b(x),\,\hp^c_d(y)]=\half\s i\s[\s\delta^c_b\hp^a_d(x)-\delta^a_d\hp^c_b(x)\s]\,\delta(x,y)\;,\no\\
  &&\hskip.1cm[\hg_{ab}(x),\,\hp^c_d(y)]=\half\s i\s[\s\delta^c_a\hg_{bd}(x)+\delta^c_b\hg_{ad}(x)\s]\,\delta(x,y)\;,\label{afc}\\
&&[\hg_{ab}(x),\,\hg_{cd}(y)]=0\;.   \no \en
These commutation relations are no more nor less than the transcription into operators of identical Poisson brackets (modulo $i\hbar$, of course) for the corresponding classical fields, namely, the spacial metric $g_{ab}(x)$ and the mixed-valence momentric field $\pi^c_d(x)\equiv \pi^{cb}(x)\s g_{bd}(x)$, along with the usual Poisson brackets between the canonical metric field $g_{ab}(x)$ and the canonical momentum field $\pi^{cd}(x)$. 

It is noteworthy that the algebra generated by $\hg_{ab}$ and $\hp^a_b$ as represented by (\ref{afc}) closes. These operators form the generators of the {\it affine group} whose elements are given by
  \bn U[\pi,\gamma]\equiv \exp[i\tint \pi^{ab}(y)\s\hg_{ab}(y)\,d^s\!y]\;\exp[-i\tint\gamma^a_b(y)\s\hp^b_a(y)\,d^s\!y]\;,\en
defined, e.g., for all smooth $c$-number functions $\pi^{ab}$ and $\gamma^a_b$ of compact
support. Since we assume that the smeared fields $\hg_{ab}$ and $\hp^a_b$ are self-adjoint
operators, it follows that $U[\pi,\gamma]$ are unitary operators for all $\pi$ and $\gamma$.

As follows directly from the affine commutation relations, we learn that
 \bn U[\pi,\gamma]^\dag\,\hg_{cd}(x)\,U[\pi,\gamma]=(\s e^{\gamma(x)/2}\s)_c^r\,{\hg}_{rs}(x)\,(\s e^{\gamma(x)/2}\s)^s_d\;, \label{w3} \;. \en
This equation establishes $\hp^c_d$ as a suitable ``conjugate'' to $\hg_{ab}$ since it is clear that such unitary transformations preserve metric positivity. 
 
\subsubsection*{Affine coherent states}
It is convenient to use affine coherent states, i.e., coherent states formed with the help of the affine group. We choose $|\eta\>$ as a normalized fiducial
vector, and we consider the set of unit vectors each of which is given by
\bn
  |\pi,\gamma\>\equiv e^{i\tint \pi^{ab}(x)\,\hg_{ab}(x)\,d^3\!x}\,e^{-i\tint\gamma^d_c(x)\,{\hat\pi}^c_d(x)\,d^3\!x}\,|\eta\>\;.  \en
As $\pi^{ab}$ and $\gamma^d_c$ range over the space of smooth functions of compact support, such vectors form a set of {\it coherent states}. The specific representation of the kinematical operators is fixed once the vector $|\eta\>$ has been chosen.

By definition, the coherent states span the original, or kinematical, Hilbert space $\frak H$,
and thus we can characterize the coherent states themselves by giving their overlap with an arbitrary coherent state.
In so doing, we choose the fiducial vector so that the overlap is 
given by
 \bn &&\hskip-.4cm\<\pi'',\gamma''|\pi',\gamma'\> \no\\
  &&\hskip0cm =\exp\bigg[-2\int b(x)\,d^3\!x\,\no\\
&&\hskip.3cm\times\ln\bigg(\frac{\det\{\half[g''^{ab}(x)+g'^{ab}(x)]+\half ib(x)^{-1}[\pi''^{ab}(x)-\pi'^{ab}(x)]\}}{\{\det[g''^{ab}(x)]\,\det[g'^{ab}(x)]\}^{1/2}}\bigg)\bigg]\;; \label{e18}\en
here $b(x)$, $0<b(x)<\infty$, is a scalar density with dimensions $L^{-3}$, that arises out of $|\eta\>$. This function is only temporary, and it will disappear when the constraints are finally enforced to their full extent.
 Several additional comments about this basic expression are in order.

Initially, regarding (\ref{e18}), we observe that $\gamma''$ and $\gamma'$ do {\it not} appear in the explicit functional form given. In particular, the smooth matrix $\gamma$ has been replaced by the smooth matrix $g$ which is defined at every point by
 \bn  g(x)\equiv e^{\gamma(x)/2}\,{\tilde g}(x)\,e^{\gamma(x)^T/2}\equiv\{g_{ab}(x)\}\;, \en
where $\gamma(x)^T$ denotes the transpose of the matrix $\gamma(x)$. Here, the positive-definite matrix ${\tilde g}(x)\equiv \{{\tilde g}_{ab}(x)\}$ is defined by the relation
   \bn {\tilde g}_{ab}(x)\equiv \<\eta|\s\hg_{ab}(x)\s|\eta\>\;. \en
Observe that the so-defined matrix $\{g_{ab}(x)\}$ is manifestly positive definite for all $x$. The map $\gamma\ra g$ is clearly many-to-one since $\gamma$
 has {\it nine} independent variables at each point while $g$, which is symmetric, has only {\it six}.
In view of this functional dependence, we may, without loss of generality, denote the given functional in (\ref{e18}) by $\<\pi'',g''|\pi',g'\>$, as well as the coherent states themselves by $|\pi,g\>$.

The diagonal coherent state matrix elements of the kinematical operators are also of interest. In particular, on the basis of the affine commutation relations, it follows that 
  \bn && \hskip.07cm\<\pi,g|\s \hg_{ab}(x)\s|\pi,g\>=g_{ab}(x)\;,  \\
   &&\hskip.14cm\<\pi,g|\s{\hat\pi}^b_a(x)\s|\pi,g\>=\pi^b_a(x) \equiv g_{ac}(x)\s\pi^{cb}(x)\;. \en
It is worthwhile emphasizing that the smooth metric $g_{ab}$ and momentric $\pi^a_b$ fields represent not {\it sharp} operator eigenvalues but, as is made clear here,  {\it mean} operator values.

\subsubsection*{Reproducing kernel Hilbert space}
The expression defined by (\ref{e18}) has the important property of being a function of positive type (often loosely called a positive-definite function).
The requirement for such a condition is that 
 \bn \sum_{k,l=1}^{N,N} \a^*_k\s\a_l\,\<\pi_k,g_k|\pi_l,g_l\>\ge 0\;, \label{t17}\en 
and this relation must hold for all variable choices, where $\{\a_k\}$  denotes a set of arbitrary complex numbers, and $N<\infty$. Besides continuity of $\<\pi'',g''|\pi',g'\>$, evident from its definition, Eq.~(\ref{t17}) is the required property for $\<\pi'',g''|\pi',g'\>$ to be a {\it reproducing kernel}, and which, therefore, can be used to define a {\it reproducing kernel Hilbert space}. 

Such a space offers a  very natural functional representation for the abstract Hilbert space ${\frak H}$ under consideration. A dense set of elements in this space is composed of functions two examples of which are given by  
  \bn && \psi(\pi,g)\equiv\sum_{k=1}^K\, \a_k\<\pi,g|\pi_k,g_k\>\;, \\  
    && \phi(\pi,g)\equiv\sum_{l=1}^L\, \beta_l\<\pi,g|\pi_{(l)},g_{(l)}\>\;. \en
The inner product of two such states is then defined by
  \bn (\psi,\phi)\equiv \sum_{k,l=1}^{K,L}\,\a^*_k\s\beta_l\<\pi_k,g_k|\pi_{(l)},g_{(l)}\>\;.  \en
Closure of this space in the norm $\|\psi\|\equiv +\sqrt{(\psi,\psi)}$ completes the reproducing kernel Hilbert space. It is important to observe from this account the important role played by the positive definite coherent state overlap function (\ref{e18}). 

\subsubsection*{Functional integral representation}
Phase space path integrals have been investigated for some time. In particular, coherent state path integrals can be used to define a very natural phase space path integral. Originally, such coherent state path integrals were defined as the continuum limit of a lattice-regularized, multi-dimensional integral. This procedure is not wrong, but the lattice itself has no close connection with the underlying continuum, especially since almost all of the underlying paths involved are discontinuous. An alternative regularization of a formal phase space path integral can be given with the help of a so-called Wiener measure regularization. In this version, true paths -- even continuous paths -- are used during the regularization and no lattice version of the time integral is required. A similar form of the functional integral, given by the limit of a Wiener measure regularized expression as the diffusion constant diverges, is valid 
in the setting of the gravitational case as well. We do not discuss the details underlying the following expression, but we  assert that (\ref{e18}) formally admits an integral representation given by
  \bn  &&\<\pi'',g''|\pi',g'\>\no\\
  &&\hskip.8cm=\exp\bigg[-2\int b(x)\,d^3\!x\,\no\\
&&\hskip1.4cm\times\ln\bigg(\frac{\det\{\half[g''^{ab}(x)+g'^{ab}(x)]+\half ib(x)^{-1}[\pi''^{ab}(x)-\pi'^{ab}(x)]\}}{\{\det[g''^{ab}(x)]\,\det[g'^{ab}(x)]\}^{1/2}}\bigg)\bigg] \no\\
&&\hskip.8cm=\lim_{\nu\ra\infty}\,{\o{\cal N}}_{\nu}\,\int \exp[-i\tint g_{ab}\s{\dot\pi}^{ab}\,d^3\!x\,dt]\no\\
  &&\hskip1.4cm\times\exp\{-(1/2\nu)\tint[b(x)^{-1}g_{ab}g_{cd}{\dot\pi}^{bc}{\dot\pi}^{da}+b(x)g^{ab}g^{cd}{\dot g}_{bc}{\dot g}_{da}]\,d^3\!x\,dt\}\no\\
&&\hskip2.3cm\times\Pi_{x,t}\,\Pi_{a\le b}\,d\pi^{ab}(x,t)\,dg_{ab}(x,t) \label{e20}\;.  \en
Here, because of the way the new independent variable $t$ appears on the right-hand side of this expression, it is natural to interpret $t$, $0\le t\le T$, $T>0$ as coordinate ``time''. The fields on the right-hand side all depend on space and time, i.e., $g_{ab}=g_{ab}(x,t)$, ${\dot g}_{ab}=\d g_{ab}(x,t)/\d t$, etc., and, importantly, the integration domain of the formal measure is strictly limited to the domain where $\{g_{ab}(x,t)\}$ is a positive-definite matrix for all $x$ and $t$. For the boundary conditions, we have $\pi'^{ab}(x)\equiv\pi^{ab}(x,0)$, $g'_{ab}(x)\equiv g_{ab}(x,0)$, as well as $\pi''^{ab}(x)\equiv\pi^{ab}(x,T)$,
$g''_{ab}(x)\equiv g_{ab}(x,T)$, for all $x$. Observe that the right-hand side holds for any $T$, $0<T<\infty$, while the left-hand and middle terms are independent of $T$ altogether.

An apparent lack of invariance of this expression under changes of the time parameterization on the right-hand side disappears in the limit that $\nu\ra\infty$.

\subsubsection*{Projection operator method for constraints}
It is well known that classical Yang-Mills theories possess constraints that form a set of closed first-class constraints. This means that the Poisson bracket of the constraint fields form (an infinite dimensional version of) a Lie algebra. Upon quantization, these constraint fields become operator fields with the property that their commutators constitute a representation of this Lie algebra, or in other words, the quantum constraints also form a set of first-class closed constraints. Gravity, on the other hand, behaves qualitatively differently. At the classical level, the Poisson brackets of the constraint fields form an open first-class system, which means that the structure constants appropriate to a closed system have become structure functions. Many of the nice properties of a first-class system still hold, such as needing to impose the constraints only once, say at time zero, and thereafter the equations of motion will ensure that the constraints are fulfilled for all time. Upon quantization of an open first-class system two things can happen, First, it may be true that the quantum constraints have the property that they still form an open first-class system of constraints. More generally, however, the quantum constraints form a (partially) second-class system, which has quite different properties. A second-class system is one for which the constraints must be continuously enforced, and it is inadequate to just enforce them at some initial time. The distinction between the two cases arises fom the nature of the structure functions. On quantization, these coefficients also become operators which may or may not commute with the constraints themselves. If these structure functions operators commute, then the first-class character of the system is preserved; if the structure function operators do not commute with the constraints, then the constraints have become partially second class. Unfortunely, this latter situation applies to quantum gravity.

The quantization of systems that are (partially) second class is traditionally
dealt with in a very different fashion than those that are strictly first class. For example, Dirac has specified that second-class constraints should be solved for at the classical level and eliminated from the theory before quantization begins. Since this can sometimes be very difficult in practice, Dirac also introduced the idea of what are now called Dirac brackets as an alternative to explictly eliminating such variables, and using these brackets in place of the standard Poisson brackets especially when promoting Poisson brackets for a basic set of kinematical variables to commutators among basic operators. Dirac brackets have the virtue that the offending second-class constraints are treated as zero classically, and that amounts to classically enforcing those constraints. In the path integral approach of Senjanovi\'c \cite{sen}, one also enforces the classical second-class constraints before quantization is introduced. However, it is well known that canonical quantization is generally valid only when the original classical canonical variables form a Cartesian set of coordinates \cite{dir}. Any quantization procedure that first enforces constraints, be they first or second class, may very well encounter the situation that the reduced phase space may not even admit Cartesian coordinates. In such cases, the suggested formalism may well give answers, but there is simply no assurance that those answers correspond to the correct results for the problem at hand. What would be preferred is a procedure to quantize a canonical system having first- and/or second-class constraints {\it before} enforcing those constraints, thereby ensuring that one has a phase space that admits Cartesian coordinates, and then enforce a reduction afterwards that can treat both first- and second-class constraints. The methods mentioned above do not have that possiblity. 

An exception to the rule of a different operator treatment for first- and second-class constraints is offered by the {\it projection operator method\/} for the quantization of systems with constraints \cite{KORIG,UNIV,SCHAL}. Let us sketch this method briefly. Rather than impose the self-adjoint quantum constraints $\Phi_\a$ in the idealized (Dirac) form $\Phi_\a\s|\psi\>_{phys}=0$, $\a\in\{1,\ldots,A\}$, on vectors $|\psi\>_{phys}$ in a putative physical Hilbert space, $|\psi\>_{phys}\in{\frak H}_{phys}$, we define a (possibly regularized) $ {\frak H}_{phys}\equiv \E \s{\frak H}$, in which $\E$ denotes a {\it projection operator} defined by
  \bn  \E=\E(\Sigma_\a\Phi_\a^2\le\delta(\hbar)^2)\;.  \en
Here $\delta(\hbar)$ is a positive {\it regularization parameter} (not a $\delta$-function!)~and we have assumed that $\Sigma_\a\Phi_\a^2$ is self adjoint. This relation means that $\E$ projects
onto the spectral range of the self-adjoint operator $\Sigma_\a\Phi_\a^2$ in the interval $[0,\delta(\hbar)^2]$.
As a final step, the parameter $\delta(\hbar)$ is reduced as much as required, and, in particular, when some second-class constraints are involved, $\delta(\hbar)$ ultimately remains strictly positive. This general procedure treats all constraints simultaneously and treats them all on an equal basis.
We can see how this procedure works by studying three simple examples \cite{PI2005}. 

First, let
$\Phi_k=J_k$, $k=1,2,3$, be the generators of the rotation group. We want to project onto
those states for which $J_k\s|\psi\>_{phys}=0$ for all $k$. We do so by considering
  \bn \E =\E(J_1^2+J_2^2+J_3^2\le \hbar^2/2)\;. \en
Since $\Sigma_k J_k^2$ is just the Casimir operator for the rotation group, with eigenvalues
given by $j(j+1)\hbar^2$, $j=0,\half,1,\dots$, it follows that $j=0$ is the only subspace
allowed by the projection operator. (Clearly, a small range of other values for $\delta(\hbar)^2$
works just as well, but we do not dwell on that aspect.)

Second, let $\Phi_1=P$ and $\Phi_2=Q$. The equations $P\s|\psi\>_{phys}=0$ and $Q\s|\psi\>_{phys}=0$
imply that $[Q,P]\s|\psi\>_{phys}=i\hbar\s|\psi\>_{phys}=0$, i.e., $|\psi\>_{phys}=0$. This is the classic example of a second-class system for which the original Dirac procedure does not work. However,
let us choose
  \bn \E=\E(P^2+Q^2\le \hbar)\;, \en
which acts to project onto vectors for which $(Q+iP)\s|\psi\>_{phys}=0$. If $Q$ and $P$ are
irreducible, then the only solution is a projection onto the ground state of an harmonic oscillator with unit angular frequency. The essential point is the projection in this case is onto a one-dimensional subspace.

It is noteworthy that the first example consists of an operator with a discrete spectrum that contains zero (first class system), while the second example involves an operator with a discrete spectrum that does {\it not} include zero (second class system).

Third, let $\Phi_1=P$ be the only constraint. This operator has its zero in the continuous
spectrum, and thus all nontrivial solutions to the equation $P\s|\psi\>_{phys}=0$ obey $\|\s|\psi\>_{phys}\s\|=\infty$. In the projection operator language, the operator
  \bn \E=\E(P^2\le \delta^2) \label{q3} \en
vanishes as $\delta\ra0$, so care must be taken to extract the ``germ" of this limit. (An $\hbar$ dependence is not important in this case.) To extract the
desired ``subspace" where ``$P=0$", it is most convenient to adopt a representation space.
For that purpose let us choose a canonical coherent state basis, where
  \bn |p,q\>\equiv e^{-iq\s P}\,e^{ip\s Q}\s|0\>\;,  \en
and where $|0\>$ is the normalized ground state of an harmonic oscillator with unit angular frequency. Then, let us focus on the quotient
\bn &&\<p'',q''|\E(P^2\le\delta^2)|p',q'\>\s{\bigg/}\s\<0|\E(P^2\le\delta^2)|0\>\\
 && \hskip.3cm=\int_\delta^\delta e^{-(k-p'')^2/2+ik(q''-q')-(k-p')^2/2}\,dk\s{\bigg/}\s\int_\delta^\delta e^{-k^2}\,dk\;.  \en
As $\delta\ra0$, the numerator and the denominator each vanish; however, the quotient will not vanish. Indeed, as $\delta\ra0$, this quotient becomes
 \bn e^{-(p''^2+p'^2)/2}\;, \label{q4} \en
which determines a reproducing kernel that characterizes a {\it one} dimensional physical Hilbert space, and which is a perfectly acceptable result in this case. Since the resultant expression no longer depends on $q''$ or $q'$, it is clear that we have reached the space where ``$P=0$". Observe that the physical Hilbert space in this case is, strictly speaking, not a subspace of the original Hilbert space ${\frak H}$.
Nevertheless, from a representation point of view, it is important to observe that the physical Hilbert space of interest can be obtained by a suitable limit taken from within the original Hilbert space ${\frak H}$.

\subsubsection*{Integral representation for the projection operator}
In special cases, such as first class systems that correspond to compact groups, it is
straightforward to find integral representations that yield an appropriate projection operator. However, it is noteworthy that there exists a universal integral representation that yields the desired projection operator for {\it any} set of constraint operators \cite{UNIV}. We have in mind the {\it operator identity} given by 
  \bn\E(\Sigma_\a\s\Phi_\a^2\le\delta(\hbar)^2)=\int {\sf T}\s e^{-i\tint_0^\tau\s\l^\a(t)\s\Phi_\a\,dt}\,\D R(\l)\;,  \en
which involves a time-ordered functional integral over $c$-number 
Lagrange multipliers, where $R(\l)$ is
a suitable (weak) measure. This result holds for any $\tau>0$ (note that the left side is independent of $\tau$). The measure $R(\l)$ depends on $\tau$, $\delta(\hbar)^2$, and the number of constraints, but it is {\it totally independent} of the choice of the set of constraint operators $\{\Phi_\a\}$. Indeed, this expression applies even if the constraint operators all {\it vanish}, in which case we learn that
  \bn 1=\int \D R(\l) \;.  \en

Such an integral representation for the projection operator can be explicitly used in forming a path integral representation for a system with constraints.
Preparatory to discussing the resultant functional integral for gravity, let us discuss the gravitational constraints themselves.

\subsubsection*{Gravitational constraints}
As is well known, there are four classical gravitational constraint functions, the three diffeomorphism constraints
  \bn  H_a(x)=-2\pi^b_{a|b}(x)\;,  \en
where ``$_|$'' denotes covariant differentiation with respect to the spacial metric $g_{ab}$, and the Hamiltonian constraint, which, in suitable units (i.e., $c^3/G=16\pi$), reads
  \bn  H(x)= g(x)^{-1/2}[\,\pi^a_b(x)\s\pi^b_a(x)-\half\pi^a_a(x)\s\pi^b_b(x)\,]+g(x)^{1/2}\s^{(3)}\!R(x)\;.
\label{h45} \en
Here $g(x)\equiv\det[g_{ab}(x)]$ and $^{(3)}\!R(x)$ denotes the scalar curvature derived from the spacial metric \cite{MTW}. Classically, these constraint functions vanish, and the region in phase space on which they vanish is called the {\it constraint hypersurface}.

It is instructive to evaluate the classical Poisson brackets among the four constraint fields. For this purpose, we enlist only the basic nonvanishing Poisson bracket given by
 \bn  \{g_{ab}(x),\pi^{cd}(y)\}=\half(\delta^c_a\delta^d_b+\delta^c_b\delta^d_a)\,\delta(x,y)\;. \en
It follows that
  \bn && \{H_a(x),H_b(y)\}=\delta_{,a}(x,y)\s H_b(x)-\delta_{,b}(x,y)\s H_a(x)\;, \label{e31}\\
    &&\hskip.13cm\{H_a(x),H(y)\}=\delta_{,a}(x,y)\s H(x)\;,  \label{e32}\\
   &&\hskip.28cm\{H(x),H(y)\}=\delta_{,a}(x,y)\s g^{ab}(x)\s H_b(x)\;.  \label{e33}\en
In these expressions, $\delta_{,a}(x,y)\equiv\d\s\delta(x,y)/\d\s x^a$, which transforms as a ``vector density''. It is clear that the Poisson brackets of the constraints vanish on the constraint hypersurface because the right-hand sides of (\ref{e31})--(\ref{e33}) all vanish there, i.e., when $H_a(x)=0=H(x)$ for all $x$. This vanishing property of the Poisson brackets is characteristic of first-class constraints.  Because of the last equation, (\ref{e33}), it is clear that the complete set of four gravitational constraint functions do not have the Poisson structure of a Lie algebra, and consequently the gravitational constraints are said to form an open first-class system of constraints. Such a situation does not automatically imply trouble in the corresponding quantum theory, but significant difficulties do arise in a number of cases. Quantum gravity is one of those cases.

Let us proceed formally in order to see the essence of the problem. Suppose that $\H_a(x)$ and $\H(x)$ represent local self-adjoint constraint operators for the gravitational field. Standard calculations lead to the commutation relations
 \bn  && [\H_a(x),\H_b(y)]=i\s[\delta_{,a}(x,y)\,\H_b(x)-\delta_{,b}(x,y)\,\H_a(x)]\;, \\
    && \hskip.18cm[\H_a(x),\H(y)]=i\s\delta_{,a}(x,y)\,\H(x)\;,  \\
 && \hskip.36cm[\H(x),\H(y)]= i\s\half\s\delta_{,a}(x,y)[\,\hg^{ab}(x)\,\H_b(x)+\H_b(x)\,\hg^{ab}(x)\,]\;,  \label{secclass}\en
where to reflect the anti-Hermitian character of the right-hand side of the last expression it is necessary for it to be properly symmetrized. In the usual Dirac approach to constraints, one asks that $\Phi_\a\s|\psi\>_{phys}=0$ for all constraints. If we assert that $\H_a(x)\s|\psi\>_{phys}=0$ and $\H(x)\s|\psi\>_{phys}=0$, then consistency holds for the first two sets of constraint commutators, but not for the third commutator in virtue of the fact that it is almost surely the case that $\hg^{ab}(x)\,|\psi\>_{phys}\not\in{\frak H}_{phys}$, even if it were smeared. As a consequence, the quantum constraints for gravity are partially second class, or as one also says, they have an anomaly.

\subsubsection*{Functional integral representation for the regularized reproducing kernel}
Combining several steps previously described, we now assert that the reproducing kernel for the regularized physical Hilbert space has the formal phase space functional integral representation given by
  \bn  && \<\pi'',g''|\s\E\s|\pi',g'\>  \no\\
  &&\hskip1cm=\int \<\pi'',g''|{\bf T}\,e^{-i\tint[\s N^a\s\H_a+N\s\H\s]\,d^3\!x\,dt}\s|\pi',g'\>\,\D R(N^a,N)\no\\
 &&\hskip1cm=\lim_{\nu\ra\infty}{\o{\cal N}}_\nu\s\int e^{-i\tint[g_{ab}{\dot\pi}^{ab}+N^aH_a+NH]\,d^3\!x\,dt}\no\\
  &&\hskip1.5cm\times\exp\{-(1/2\nu)\tint[b(x)^{-1}g_{ab}g_{cd}{\dot\pi}^{bc}{\dot\pi}^{da}+b(x)g^{ab}g^{cd}{\dot g}_{bc}{\dot g}_{da}]\,d^3\!x\,dt\}\no\\
  &&\hskip2cm\times\bigg[\Pi_{x,t}\,\Pi_{a\le b}\,d\pi^{ab}(x,t)\,dg_{ab}(x,t)\bigg]\,\D R(N^a,N)\;. \label{f39}\en
Here we have introduced a Wiener measure regularization as a devise in order to greatly help in the definition of this functional integral. In view of the infinitely many spacial degrees of freedom, an additional regularization is needed to handle that aspect, but the Wiener measure regularization makes a lattice treatment of the time integration unnecessary.

In this last expression we have reached an important formal relation. Despite the general canonical appearance of ({\ref{f39}), we emphasize once again that this representation has been based on the affine commutation relations and {\it not} on any canonical commutation relations! 

Of course, (\ref{f39}) requires several regularization parameters to be well defined: one because it is a field theory, and another because we are only enforcing the constraints in a regularized manner. However, just as the elementary example above demonstrated [cf., Eqs.~(\ref{q3})-(\ref{q4})], as we remove these regulariztion parameters, we expect that (\ref{f39}) will pass to a continuous, positive-definite function that defines a reproducing kernel for the physical Hilbert space even if that space was not a literal subspace of the original Hilbert space; cf.,  \cite{kl4}.

\subsubsection*{Hard core picture of nonrenormalizability}
In very general terms, a functional integral appears as
 \bn Z_\l(h)={\cal M}\int e^{i\s[\tint hf +Q(f)+\l NQ(f)\,]}\,\D f\;,  \en
where $h$ denotes a source field, $f$ denotes fields which are to be integrated, $Q(f)$ is a quadratic combination of the fields, and $NQ(f)$ is a nonquadratic function of the fields representing the nonlinear interactions that are present. The parameter $\l\ge0$ is a coupling constant. As usual ${\cal M}$ is chosen so that $Z_\l(0)=1$ for all $\l$. If $\lim_{\l\ra0}\s Z_\l(h)=Z_0(h)$, the free theory, we say that $NQ$ represents a {\it continuous perturbation}; if $\lim_{\l\ra0}\s Z_\l(h)\not=Z_0(h)$, we say that $NQ$ represents a {\it discontinuous perturbation}. It would seem self evident that all interactions should be continuous, and, most importantly, this is the assumption underlying perturbation theory. However, there are examples of discontinuous perturbations as well, and it can be argued that the so-called nonrenormalizable theories fit into this category. How can a discontinuous interaction arise? 

Formally, if $|NQ(f)|=\infty$ for a set of nonzero measure for which $|Q(f)|<\infty$, then we say that the interaction acts partially as a {\it hard core}, projecting out certain paths that would otherwise be allowed were the interaction absent altogether. These paths are projected out for all $\l>0$ no matter how small, and thus, when $\l\ra0$, the paths remain absent and so the expected limit $Z_0(h)$ does not occur. Instead, the limit becomes (say) $Z'_0(h)$, the pseudofree theory, an expression that retains the irremovable effects of the hard core.  
The moral is: Once turned on, such interactions cannot be completely turned off! Evidently, the presence of a hard-core interaction makes any perturbation theory developed about the original, unperturbed theory almost totally meaningless. However, since the interacting theory is {\it continuously connected} to the pseudofree theory, it may well possess some form of perturbation theory about the pseudofree theory. 
   
The hard-core picture can be forcefully argued for scalar fields with quartic interactions, the so-called $\varphi^4_n$ models, where $n\ge5$ denotes the number of space time dimensions; see \cite{kbook}. Arguments can be advanced for scalar fields that a novel counterterm is required which is not one suggested by perturabtion theory. In particular, the form of the counterterm is believed to be proportional to $\hbar^2/\varphi^2$, a term that scales as the kinetic energy. 

\subsubsection*{Nonrenormalizable gravity}
Although the differences between gravity and nonrenormalizable scalar interaction are significant in their details, there are certain similarities we wish to draw on. Most importantly, one can argue \cite{nongrav} that the nonlinear contributions to gravity act as a hard-core interaction in a quantization scheme, and thus the general picture sketched above for nonrenormalizable fields should apply to gravity as well. Assuming that the analogy holds further, there should be a nonstandard, nonclassical counterterm that incorporates the dominant, irremovable effects of the hard-core interaction. Accepting the principle that in such cases perturbation theory offers no clear hint as to what counterterms should be chosen, we appeal to the guide used in the scalar case. Thus, as our proposed counterterm, we look for an ultralocal potential that scales as the kinetic energy which appears in the Hamiltonian constraint. In fact, the only ultralocal potential that has the right transformation properties is proportional to $\hbar^2\s g(x)^{1/2}$ \cite{kl3}. Thus we are led to conjecture that the ``nonstandard counterterm'' is none other than a term like the familiar cosmological constant contribution! Indeed, the role of the cosmological constant term has been discussed in other contexts as well; cf., \cite{reu}.

\end{document}